# High-order localized spoof surface plasmon resonances and experimental verifications

Zhen Liao[1], Yu Luo[2, 3], Antonio I. Fernández-Domínguez[3, 4], Xiaopeng Shen[1], Stefan A. Maier[3], Tie Jun Cui[1,5*]

[1] *State Key Laboratory of Millimetre Waves, Southeast University, Nanjing 210096, China*

[2] *Photonic Centre of Excellence (OPTIMUS), School of Electrical and Electronic Engineering, Nanyang Technological University, Nanyang Avenue 639798, Singapore*

[3] *The Blackett Laboratory, Department of Physics, Imperial College London, London SW7 2AZ, UK*

[4] *Departamento de Fisica Teorica de la Materia Condensada, Universidad Autonoma de Madrid, E-28049 Madrid, Spain*

[5] *Cooperative Innovation Centre of Terahertz Science, No.4, Section 2, North Jianshe Road, 610054 Chengdu, China*

[*] *Email: tjcui@seu.edu.cn*

We theoretically demonstrated and experimentally verified high-order radial spoof localized surface plasmon resonances supported by textured metal particles. Through an effective medium theory and exact numerical simulations, we show the emergence of these geometrically-originated electromagnetic modes at microwave frequencies. The occurrence of high-order radial spoof plasmon resonances is experimentally verified in ultrathin disks. Their spectral and near-field properties are characterized experimentally, showing an excellent agreement with theoretical predictions. Our findings shed light into the nature of spoof localized surface plasmons, and open the way to the design of broadband plasmonic devices able to operate at very different frequency regimes.

Surface plasmons (SPs), coherent electronic oscillations at the interface of materials with permittivities of opposite sign (typically metals and dielectrics), have been intensively investigated in recent years [1, 2]. At optical and infrared frequencies, light can couple to SPs and create propagating electromagnetic (EM) waves confined at metal-dielectric interfaces, the so-called surface plasmon polaritons (SPPs) [3]. There is another family of plasmonic modes, termed localized surface plasmons (LSPs), which originate from the optical excitation of SPs in finite metallic nanoparticles [4]. At LSP resonance, EM field amplification occurs both inside and in the near-field of the nanoparticle, leading as well to a significant increase of its extinction cross section [4]. Taking advantage of these features, LSPs have been exploited in a broad range of technological areas, such as chemical and biological sensing [5, 6], Raman sensors [7], and plasmonic antennas [8, 9]. When plasmonic nanoparticles are much smaller than the wavelength of light, the quasi-static approximation is valid and only dipolar LSP modes are active. However, when for increasing particle size, multipolar resonances emerge [10-16].

At regimes of the EM spectrum far lower than the plasma frequency (such as far infrared, terahertz, and microwave frequencies), metals behave as perfect conductors, which prevents them from supporting SP modes. To transfer the capabilities of SPPs to these frequency ranges, spoof (or designer) surface plasmons have been proposed [17-20]. These geometrically-originated modes mimic the dispersion characteristics and confinement properties of SPPs in perfect conducting surfaces structured at a sub-wavelength scale. The spoof plasmon concept has made possible the design and realization of SPP-components [21-25] operating at low spectral ranges. Lately, corrugated metallic strips with nearly zero thickness have been shown to support highly-confined conformal SPP waves, which open the way to the design of compact, ultrathin, microwave and terahertz plasmonic functional devices [26-29]. Unlike spoof SPPs, which have been widely studied in the past decade, spoof LSPs have been theoretically proposed [30] and experimentally tested only very recently [31]. Specifically, the occurrence of microwave multipolar spoof LSPs was demonstrated in metallic cylinders and disks decorated by periodic one-dimensional (1D) groove arrays.

The analogy between conventional SP modes at optical frequencies and low-frequency spoof plasmons is not complete. Contrary to two-dimensional (2D) corrugations (such as holes or dimples), the spoof plasmon fields inside 1D apertures (such as grooves or slits) is always propagating [17, 32]. This establishes a fundamental difference between spoof and conventional SPs, whose tail inside the supporting metal is always evanescent. Importantly, although 1D spoof plasmon structures were firstly associated with positive effective permittivity parameters [33], it has been shown recently that there are also significant discrepancies between spoof SP and dielectric resonators [34]. There exist, though, a common feature shared by dielectric and spoof SPP waveguides. Both support high-order modes, associated with Fabry-Perot-like cavity resonances within the plane normal to the direction of propagation. Several theoretical studies have predicted the existence of these high-order spoof SPPs [32-33, 35-36], which have been also recently reported experimentally [37].

In this article, we introduce and verify experimentally the emergence of high-order radial spoof LSPs in periodically corrugated metal particles. These modes resemble the optical whispering gallery modes [38] sustained by dielectric resonators, whose high Q-factor characteristics are currently being exploited in a wide range of technologies [39, 40]. We provide analytical and numerical insights into the emergence of these high-order radial resonances for different azimuthal dependencies (hexa-, octo-, deca-, dodeca-, tetradeca-, and hexadeca-pole modes). Finally, we design, fabricate and characterize an ultrathin corrugated mm-sized spoof plasmon disk. Our microwave experiments demonstrate that this structure supports second-order radial octopole and decapole LSPs.

## Results

**Theory and analysis.** First, we investigate theoretically the LSPs supported by the textured 2D perfectly conducting cylinder (with radius $R$) shown in Fig. 1 [30, 41]. The structure consists of an inner core of radius $r$ surrounded by periodic array of grooves of pitch $d = 2\pi R / N$ (where $N$ is the number of grooves), height $h=R\text{-}r$, and width $a$. For simplicity, we assume that the dielectric constant filling grooves is $n = 1$. The system is excited under TM plane wave illumination. Using Mie scattering and an effective medium approach (see the method section), an analytical expression for the scattering cross section (SCS) of the corrugated cylinder can be expressed as $\sigma = \dfrac{4}{k_0} \sum\limits_{n=-\infty}^{n=\infty} |b_n|^2$. The spoof LSP resonances supported by the structure are given by the singularities in the Mie expansion coefficients, $b_n$ (where $n$ labels the azimuthal characteristics of the LSP mode). The vanishing condition for the denominator of the $n$-coefficient can be expressed as

$$\frac{a}{d} \frac{J_1(k_0 r)Y_1(k_0 R) - J_1(k_0 R)Y_1(k_0 r)}{J_1(k_0 r)Y_0(k_0 R) - J_0(k_0 R)Y_1(k_0 r)} = \frac{H_{n+1}^{(1)}(k_0 R) - H_{n-1}^{(1)}(k_0 R)}{2H_n^{(1)}(k_0 R)}. \qquad (1)$$

where $J_n()$ and $Y_n()$ are Bessel functions of the first and second kind, respectively and $H_n^{(1)}() = J_n() + iY_n()$.

We can gain physical insight into Equation (1) by taking the limit $k_0 R, k_0 r >> 1$, in which the asymptotic form of the Bessel functions can be used. In this limit, Equation (1) can be rewritten as

$$\frac{n}{R} = k_0 \sqrt{1 + \left(\frac{a}{d}\right)^2 \tan^2\left(k_0(R-r)\right)}, \tag{2}$$

which identifying the right hand side with the propagating wavevector, $k = n/R$ (note that the azimuthal index $n$ gives the number of modal wavelengths which fit within the cylinder perimeter), recovers the spoof SPP dispersion relation in a 1D array of grooves [17]. Despite their approximate character (the effective medium approximation is only valid in the limit $a, d << \lambda$), Equation (2) establishes a clear analogy between spoof SPs in corrugated flat surfaces and spoof LSPs in textured cylinders. Note that the structures we consider here do not fulfil the conditions which led to Equation (2). However, an equivalent relation can be obtained from Equation (1) in the most realistic configuration, $k_0 R >> 1, k_0 r << 1$ (small core radius), having

$$\frac{n}{R} = k_0 \sqrt{1 + \left(\frac{a}{d}\right)^2 \frac{\tan\left(k_0 R - \frac{\pi}{4}\right) - \frac{\pi}{4}(k_0 r)^2}{1 + \frac{\pi}{4}(k_0 r)^2 \tan\left(k_0 R - \frac{\pi}{4}\right)}} \tag{3}$$

Figure 1a depicts the scattering cross section (SCS) for a corrugated cylinder of outer radius $R = 25$ mm versus the incident frequency and $h/R = (R-r)/R$ for a groove width $a = 0.5d$. Note that the cross sections are normalized to the structure physical cross section. The contour plot was calculated using the effective medium approach described above, and the blue dots render the resonant condition obtained from Equation (3). The spectra show a prominent scattering maximum band which redshifts with increasing groove depth. This band comprises several LSP resonances [30, 41], as reflected by our analytical predictions. Importantly, Fig. 1a shows another set of SCS peaks for large groove depths and high frequencies (green dots). The effective medium field maps evaluated at the various SCS maxima for $r = 2$ mm are rendered in Fig. 2. These results show that the modes within each scattering band share the same radial characteristics, but present different azimuthal properties. Importantly, whereas the LSPs in the first band have been investigated in the past [30,41], to the best of our knowledge, the high frequency resonances, whose fields exhibit a node along the groove depth, had not been observed yet. It is the aim of this work to characterize in detail these high, second order radial modes.

Figure 1a indicates that in order to design a spoof SP resonator supporting high-order radial LSPs at lower frequencies, deep grooves must be carved at the cylinder surface. Thus, we fix the outer and inner radii to $R = 25$ mm and $r = 2$ mm. Fig. 1b illustrates how, within the effective medium approximation, the ratio $a/d$ affects slightly the scattering properties of the structure, yielding a small redshift of the LSP resonances with increasing groove width. In the following, we set $a/d = 0.5$ for a feasible, three-dimensional full electromagnetics numerical design and characterization of our spoof LSP resonators.

In order to retain the EM properties of the previous 2D textured cylinders, we consider thick disks with a thickness-to-radius aspect ratio $L/R = 10$. The extinction cross section (ECS) of a representative textured cylinder, simulated using the full-wave commercial software CST Microwave Studio, is shown in Fig. 3 (black solid line). The parameters of the structure are in which the parameters are chosen as $R = 25$mm, $r = 2$ mm, $L = 250$ mm, $N = 60$ ($d = 2.6$ mm), and $a = 0.5d = 1.3$ mm (see inset of Fig. 3). From the simulation results, we observe that the 3D ECS contains two well-separated bands as well, both of which have multiple peaks. As discussed above, we expect the lower (higher) extinction band to encompass spoof LSP modes presenting zero (one) field nodes along the radial direction inside the grooves. The first (second) set of modes are labelled as 1 to 5 (6-9). The SCS obtained from our effective mode theory for a textured 2D cylinder with the same geometric parameters is plotted in red dashed line. Note that whereas the theoretical SCS has been normalized to the 2D cylinder diameter $2R$, the numerical ECS value has been normalized to the disk side area $2RL$. Remarkably, both normalized spectra are in excellent agreement (note that the absorption cross section of both structures vanishes, as we are using a perfect conducting model for the metal permittivity).

To further confirm the nature of the spoof LSP modes governing the EM response of 3D our textured metal disk, Fig. 4 shows the near electric field patterns for the 9 cross section maxima identified in Fig. 3. Specifically, the $z$-component of the electric field is plotted within the $x$-$y$ plane 1.5 mm above the disk upper surface. In all panels, the colour scale has been saturated to show clearly the mode profile, ranging from red (positive) to blue (negative). Fig. 4 demonstrates that the corrugated disks supports two kinds of LSP modes. The first five LSPs (panels a-e) do not present radial nodes, and correspond to fundamental dipole to decapole modes. These were evaluated at the frequencies labelled as 1 to 5 in Fig. 3, and have been reported earlier in theory [30] and experiments [31]. Importantly, the dipole (a) and quadrupole (b) LSPs are resonances are not evident in the ECS spectrum (see Fig. 3), which is consistent to what observed in Ref. [30]. Equation (3) sheds light into this observation. It is straightforward to show that it cannot be satisfied for $n = 1$ (dipole) and $n = 2$ (quadrupole) for the geometric parameters of the structure. Thus, the spectral shoulders labelled as 1 and 2 in Fig. 2 emerge due to a maximum, rather than a divergence, in the corresponding Mie scattering coefficient.

The last four electric field maps in Fig. 4 (panels f-i) correspond to the cross section maxima 5 to 9 in Fig. 3 and show azimuthal characteristics corresponding to dodeca to octadecapole modes. However, all these field profiles present a radial node along the grooves, and can be identified as second-order radial spoof LSPs. It is noteworthy that all these patterns are in accordance with the field distributions based on effective medium theoretical results shown in Fig. 2. Whereas the first set of resonances could be related to the LSPs sustained by metallic nanoparticles, the emergence of these high-order radial modes indicates that the analogy with optical whispering gallery modes in dielectric resonators [38] is more appropriate.

Recently, it has been reported that fundamental spoof LSP resonances (whose electric field does not exhibit radial nodes) are significantly altered when the disk thickness is reduced to values much smaller than $R$ [31]. In the ultrathin limit, the dipole and quadrupole resonances became clearly apparent (note that this cannot be reproduced by Equation (3), as it is valid only for large $L$ ). In Fig. 5, the dependence of second-order radial spoof LSPs on the disk thickness is analysed. This figure shows that, for the geometric parameters considered in Fig. 3, reducing the disk thickness from $L$=25 mm to $L$=0.018 mm the ECS maxima experience a moderate red-shift, while the most apparent resonance shifts from a second-order radial dodeca to a decapole LSP mode. This observation can be understood through the analogy that Equation (2) establishes between spoof LSPs and spoof SPPs [26] (see Fig. 6). Besides, reducing the disk thickness also results in an increase of the quality factor for each LSP resonance, thereby rendering the low frequency modes (originally absent for large disk thickness) experimentally observable.

Figure 7 shows the ECS spectrum between 7.5 and 10 GHz for an ultrathin corrugated metallic disk with nearly zero thickness (see inset), with geometric parameters $R$ =25mm, $r$ =2mm, $L$ =0.018mm, $N$ =60, $d$=2.61 mm, and $a$ =0.5 $d$ . The emergence of second-order radial spoof LSP resonances translates into various extinction maxima. In order to reveal the nature of the LSPs, we simulate the near electric fields at the frequencies labelled as 1 to 6, and shown in the right panels of Fig. 7. We clearly observe second-order radial hexa-, octo-, deca-, dodeca-, tetradeca-, and hexadecapole spoof LSP modes. Note that the lowest second-order radial modes (hexa- and octopole), labelled as 1 and 2, are not apparent in the extinction spectrum, in a similar way as it was observed for thick disks in Figs. 3 and 4. Fig. 7 shows that, as we already mentioned, the most intense spoof second-order radial LSP in thin disks is a decapole mode, whereas it was a dodecapole mode in the case of thick cylinders.

**Experimental results.** We fabricate an ultrathin ( $L$ =0.018 mm) corrugated cooper disk for the sake of experimental verification of the second-order radial LSPs predicted by theory and numerical calculations. The structure is shown in the Fig. 8a, whose geometric parameters are chosen to be the same as in Fig. 7 ( $r$ =2mm, $R$ =25mm, $L$ =0.018mm, $N$ =60, $a$ =0.5 $d$ ). The cooper disk is etched on a 0.2mm-thick substrate with dielectric constant 3.5 and loss tangent 0.02. In order to investigate the influence of the dielectric substrate on the high-order radial spoof LSPs supported by the system, we performed CST Microwave Studio simulations in which both cooper and dielectric losses were considered to match the experimental conditions. The system is excited using a linearly polarized plane wave. The numerical ECS spectrum is plotted in black solid line in Fig. 8a. The comparison with Fig. 7 (without substrate) shows that the effect of the dielectric substrate is a slight redshifting of the spoof LSP frequencies. On the other hand, the introduction of dielectric and metal losses translates into the degradation of the spoof LSPs quality factor. Thus, the dominant, low frequency, resonance peaks (1-3 in Fig. 7) have been broaden, and the sharp, high frequency modes (4-6 in Fig. 7) have vanished. Fig. 8b render the simulated near electric-field patterns at the frequencies indicated in panel (a). As expected, the modes 1 and 2 can be clearly identified with second-order radial octopole and decapole spoof LSP resonances, respectively.

To demonstrate the predicted ECS spectrum experimentally, we measured the near field response of the fabricated LSP structure, as presented in Fig. 8a (the red line). Fig. 8a shows the excellent agreement between experiment and simulations, and the measured spectrum exhibits extinction peaks at 7.75 GHz and 7.96 GHz very close to the numerical spoof LSP resonances.

Figure 8c renders the measured near-field maps at 7.75 GHz and 7.96 GHz, which are the resonant frequencies identified from the measured spectrum. These maps are in very good agreement with the full-wave simulations in panel (b), which demonstrates experimentally the emergence of second-order radial spoof octopole and decapole LSPs in our corrugated cooper disk sample.

## Conclusion

In summary, we have introduced the concept of high-order radial spoof localized surface plasmon resonances. Using an effective medium theory, we have provided analytical insight into the properties of these electromagnetic modes in corrugated perfect metal cylinders. Through numerical simulations, we have showed their emergence in thick and ultrathin disks at microwave frequencies. Finally, we have verified experimentally the occurrence of second-order radial octo- and decapole spoof plasmon resonances, showing an excellent agreement between numerical predictions and measurements for both spectral and near-field characteristics. Our findings shed light into the actual nature of spoof localized surface plasmons, and open the way to the design of plasmonic broadband devices, which exploiting high-order radial resonances, can be able to operate efficiently at very different frequency ranges (from the microwave to the terahertz regime).

## Methods

**Experimental setup and measurement.** The near electric-field response of the spoof LSP sample is measured by means of two monopole antennas. First, a transmitter antenna placed 3 mm away from the disk plane is used to excite the spoof LSP modes. Second, a receiving antenna, able to move freely within the plane 1.5 mm above the sample, is used to probe the near-field response of the system, as illustrated in Fig. 9. Both monopole antennae are connected to an Agilent vector network analyser.

To map the near spoof LSP electric-field distributions experimentally, we make use a metamaterial-based Vivaldi antenna [42] to produce quasi-plane waves in the ultrathin corrugated disk plane. A coaxial probe is used to record the z-component of the electric field (out of the disk plane) at different locations within a plane 1.5 mm above the sample surface. The experimental setup is shown in Fig. 10.

**Theoretical derivations based on Mie-Lorentz theory.** Within the effective medium approximation, and using Mie scattering theory, the fields inside and outside the cylinders can be written as

$$H_z = \begin{cases} \sum_{n=-\infty}^{\infty} \left[ i^n J_n\left(k_0\rho\right) + b_n H_n^{(1)}\left(k_0\rho\right) \right] e^{in\varphi}, & \rho \geq R \\ \sum_{n=-\infty}^{\infty} \left[ c_n J_0\left(k_0\rho\right) + d_n Y_0\left(k_0\rho\right) \right] e^{in\varphi}, & R \geq \rho \geq r \end{cases}$$

where $J_n$, $Y_n$, and $H_n^{(1)}$ are Bessel function, Neumann function, and Hankel function of the first kind. The unknown coefficients can be obtained through the following matching equations at the interface $\rho = R$:

$$i^n J_n\left(k_0 R\right) + b_n H_n^{(1)}\left(k_0 R\right) = c_n J_0\left(k_0 R\right) + d_n Y_0\left(k_0 R\right)$$

$$i^n J_n'\left(k_0 R\right) + b_n H_n^{(1)\prime}\left(k_0 R\right) = -\frac{a}{d}\left[ c_n J_1\left(k_0 R\right) + d_n Y_1\left(k_0 R\right) \right]$$

$$c_n J_1\left(k_0 r\right) + d_n Y_1\left(k_0 r\right) = 0$$

from which, we obtain

$$b_n = -i^n \frac{\dfrac{a}{d} J_n\left(k_0 R\right) f - J_n'\left(k_0 R\right) g}{\dfrac{a}{d} H_n^{(1)}\left(k_0 R\right) f - H_n^{(1)\prime}\left(k_0 R\right) g}$$

$$c_n = \left[ i^n J_n\left(k_0 R\right) + b_n H_n^{(1)}\left(k_0 R\right) \right] Y_1\left(k_0 r\right)$$

$$d_n = -c_n \frac{J_1\left(k_0 r\right)}{Y_1\left(k_0 r\right)}$$

where *f* and *g* are defined as:

$$f = J_1(k_0 r) Y_1(k_0 R) - J_2(k_0 R) Y_1(k_0 r)$$
$$g = J_0(k_0 R) Y_1(k_0 r) - J_1(k_0 r) Y_0(k_0 R)$$

## Acknowledgments

Z. L. and Y. L. contribute equally in this work. This work was supported in part by the National Science Foundation of China (60990320, 60990321, 60990324, 61171024, 61171026, and 61138001), in part by the National High Tech (863) Projects (2012AA030402 and 2011AA010202), in part by the 111 Project (111-2-05), in part by the Scientific Research Foundation of Graduate School of Southeast University (Grant No. YBJJ1436), in part by the Program for Postgraduates Research Innovation in University of Jiangsu Province (Grant No. 3204004910) and in part by the Fundamental Research Funds for the Central Universities. Y. L. and S. A. M. are supported by the Leverhulme trust. Y. L. would like to acknowledge the funding support from NTU-A*STAR Silicon Technologies Centre of Excellence under the program grant No. 11235150003.

## Author contribution statement

Z. L. and Y. L. contribute equally in this work. Z. L., X. S., and T. J. C. conducted the experiment, Y. L., A. I. F. D., and S. A. M. performed the theoretical calculation, Z. L., A. I. F. D., and T. J. C. wrote the manuscript, all the authors discussed the results and edited the manuscript.

## Additional information

The author declares no competing financial interests.

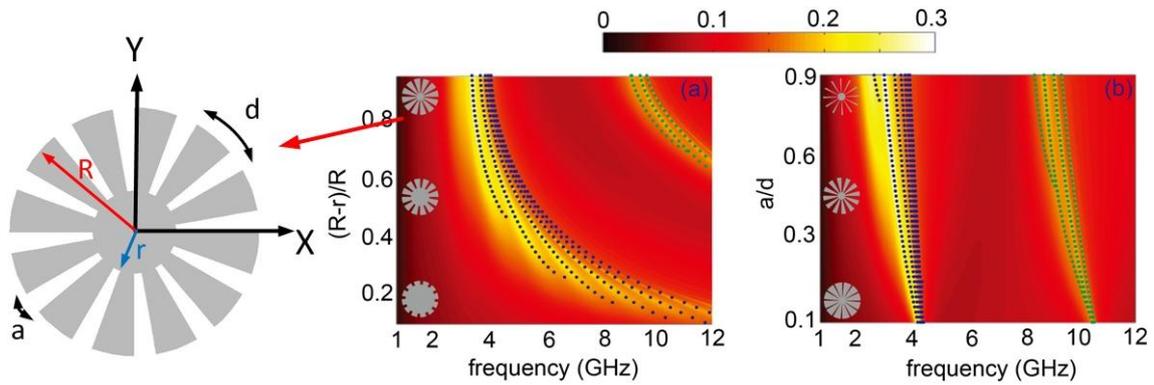

**Figure 1** (a) Normalized scattering cross section of textured perfect conducting 2D cylinders as a function of the frequency and the ratio $(R-r)/R$. The geometrical parameters are set as $R = 25$ mm and $a = 0.5d$. (b) Normalized scattering cross section as function of the frequency and $a/d$. The geometrical parameters are set as $R = 25$ nm and $r = 2$ nm. The left sketch and insets show the geometric parameters of the structure. The blue and green dots are plotted using Eq. (3)

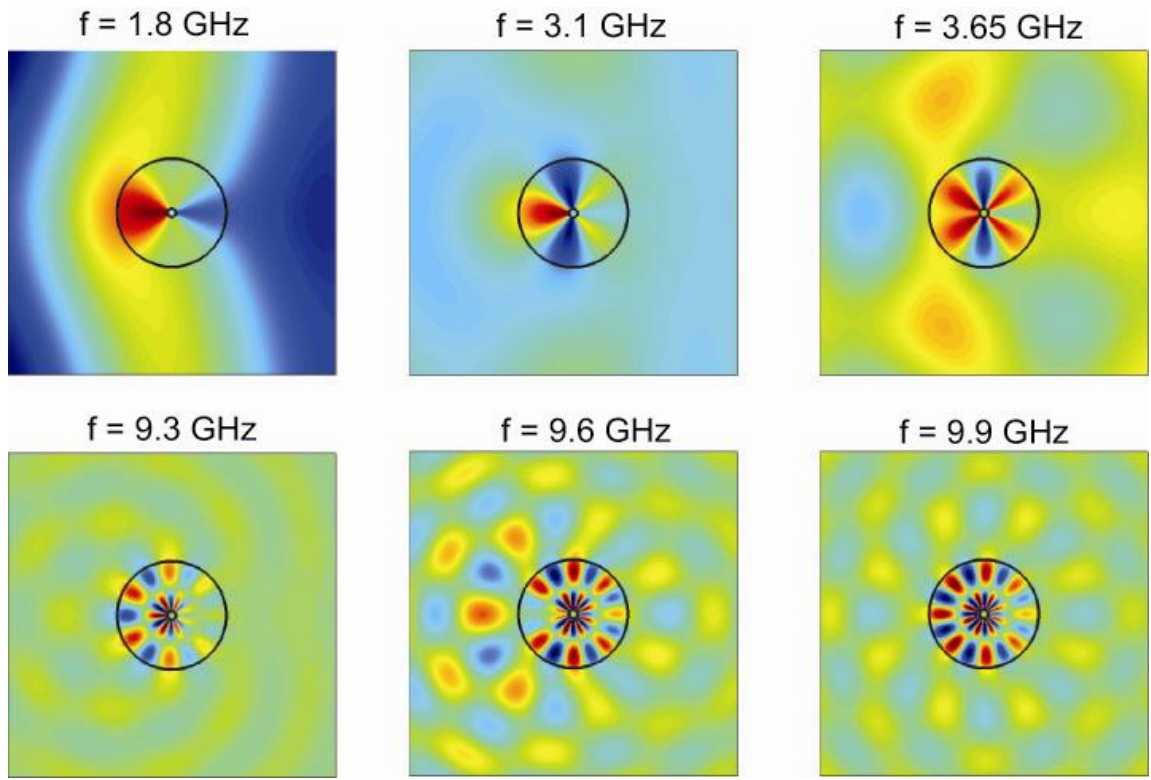

**Figure 2** Effective medium field distributions evaluated at different frequencies for a corrugated perfect conducting cylinder with $R = 25$ mm, $r = 2$ mm, and $a = 0.5d$.

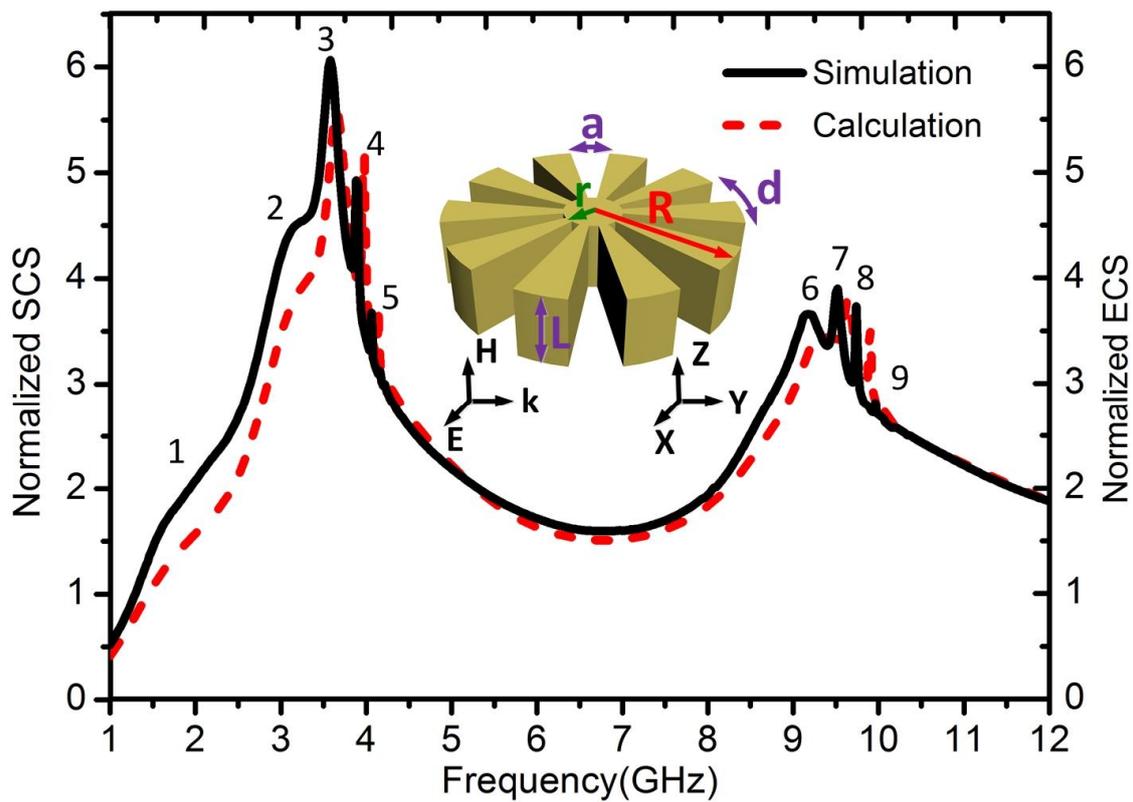

**Figure 3** Simulated ECS (black solid line) and theoretically calculated SCS (red dashed line) for a corrugated metallic disk (see inset) with $R$ =25 mm, $r$ =2 mm, $L$ =250 mm (note that the structure is 2D in the theoretical calculations), $N$ =60 ( $d$ = 2.6 mm), and $a = 0.5d = 1.3$ mm. The numbers indicate the LSP resonance ordering.

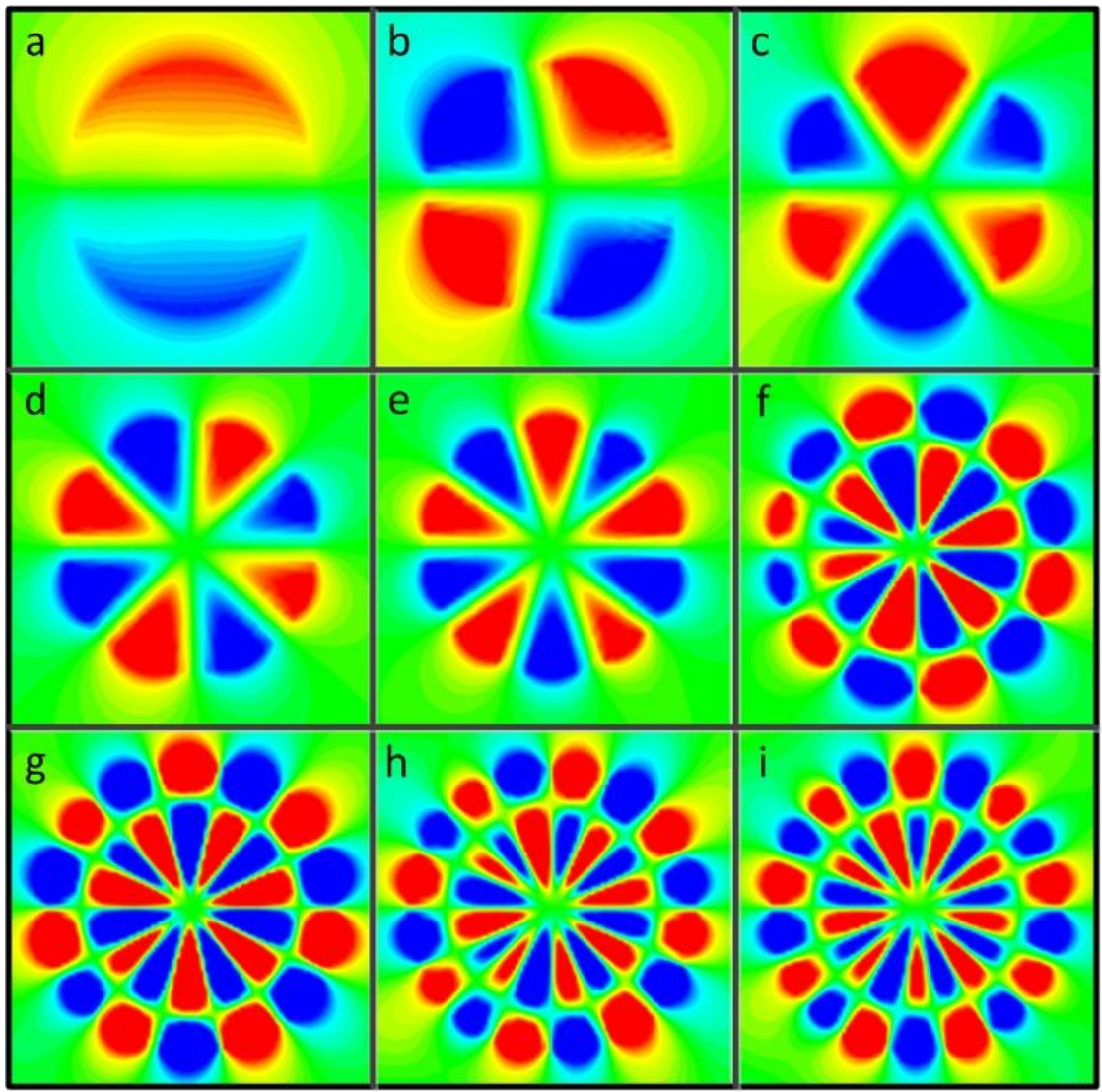

**Figure 4** Simulated near-field maps showing the *z*-component of the electric field within the *x-y* plane 1.5 mm above the upper surface of the corrugated metallic disk considered in Fig. 2. Panels a to e (f to i) show the fundamental (second-order) spoof LSPs labelled as 1 to 5 (6 to 9) in Fig. 3. The colour scales are saturated in each case to make the field profiles more apparent.

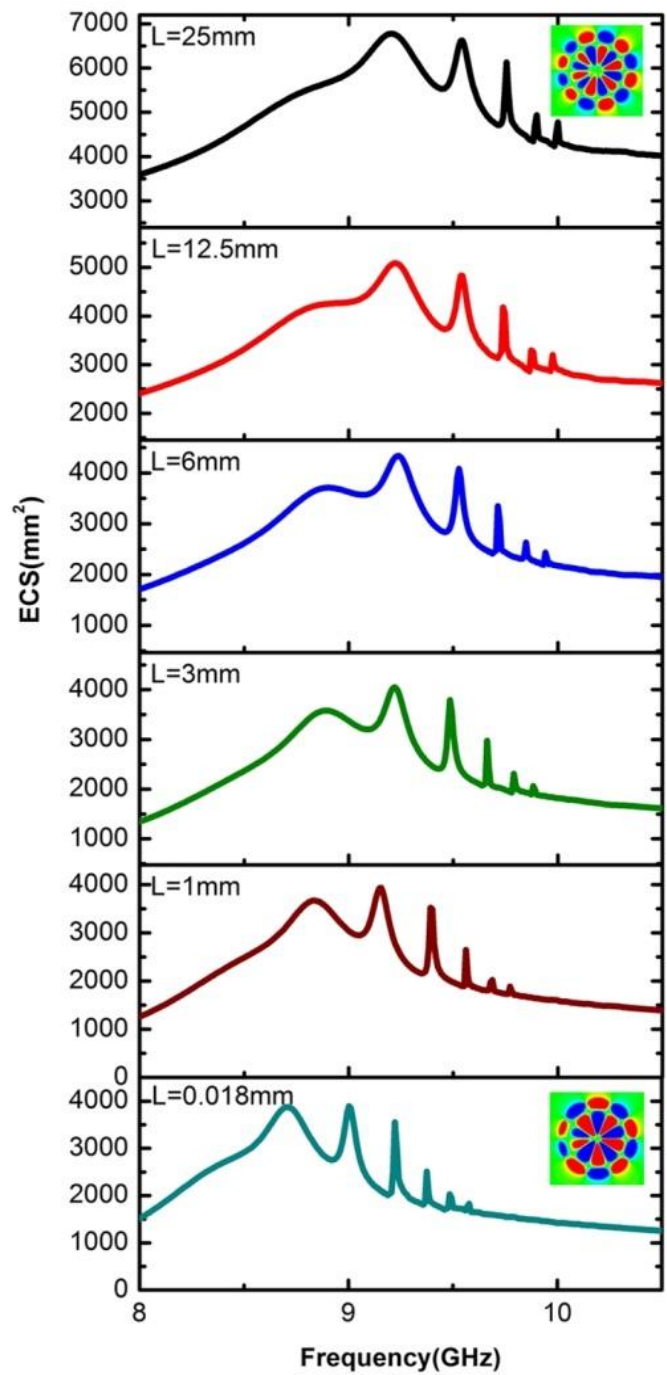

**Figure 5** Simulated ECS spectra (between 8 and 11 GHz) showing the signature of various second-order spoof LSPs for *L*=25, 12.5, 6, 3 and 0.018 mm (panels from top to bottom. The insets in the top and bottom panels render the most intense resonance: dodecapole (decapole) LSP for *L*=25 mm (*L*=0.018 mm).

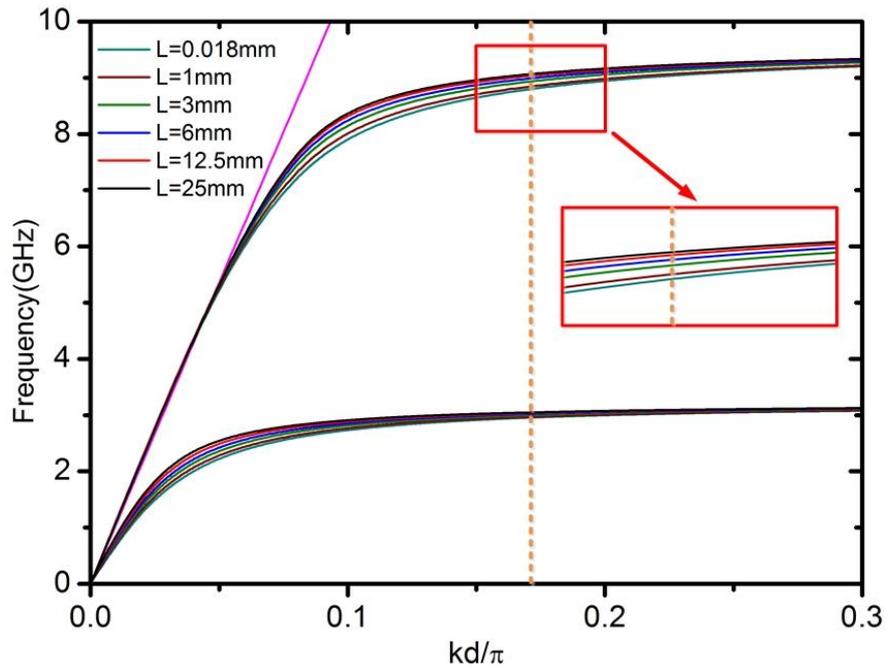

**Figure 6** Spoof SPP dispersion bands for groove arrays of different thicknesses, *L*, and *a=0.5d, h=*23mm ,and *d=1.4* mm (parameters equivalent to the texture in Fig. 5). Two different set of modes can be observed, each of which can be associated with a different divergences for the tangent function in Eq. (2). The lower (upper) band is linked to the resonant condition $k_0 h = \pi/2$ ( $k_0 h = 3\pi/2$ ), and correspond to the fundamental (second order) spoof SPP mode. The orange dashed line shows the resonance condition ( $k = n/R$ ) for the decapole resonance in Fig. 5. The inset shows an enlargement of the second-order dispersion, which explains qualitatively the LSP red-shifting with decreasing *L* observed in Fig. 5.

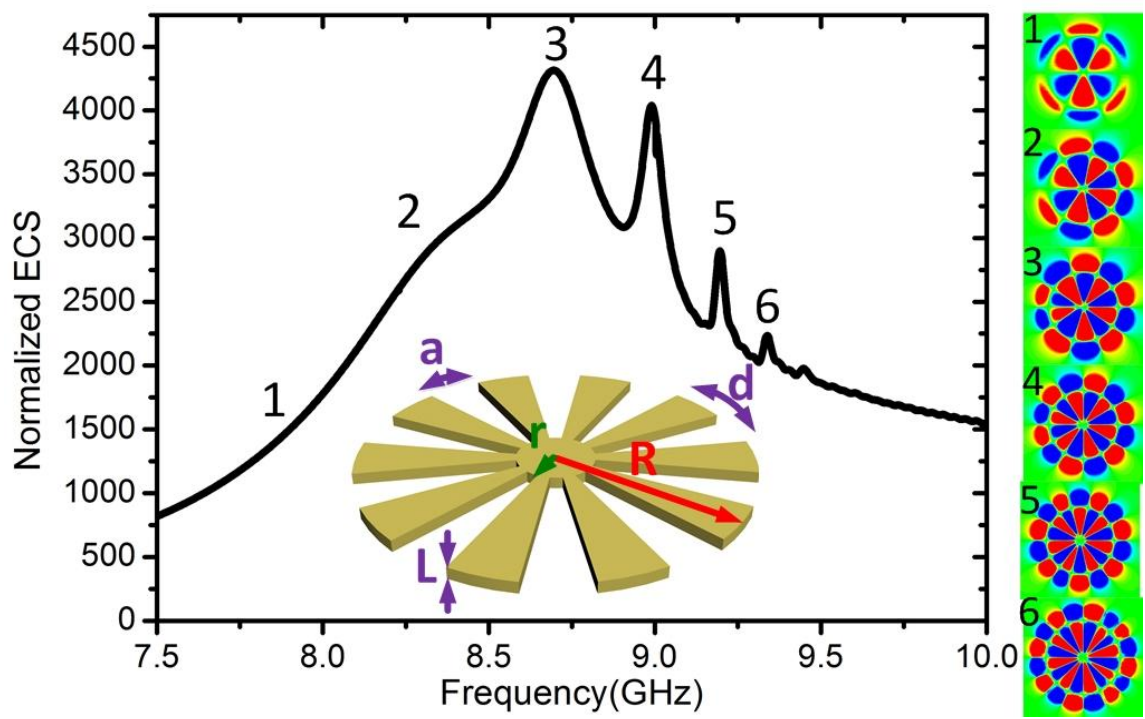

**Figure 7** Normalized ECS spectrum for an ultrathin corrugated metallic disk with $R$ =25mm, $r$ =2mm, $L$ =0.018mm, $N$ =60, $d$=2.61mm, and $a$ =0.5$d$ . The right panels display the near electric-field distributions on the *x-y* plane which is 1.5 mm above the upper surface of the disk at the resonant frequencies, corresponding to the second-order hexa-, octo-, deca-, dodeca-, tetradeca-, and hexadecapole modes, respectively.

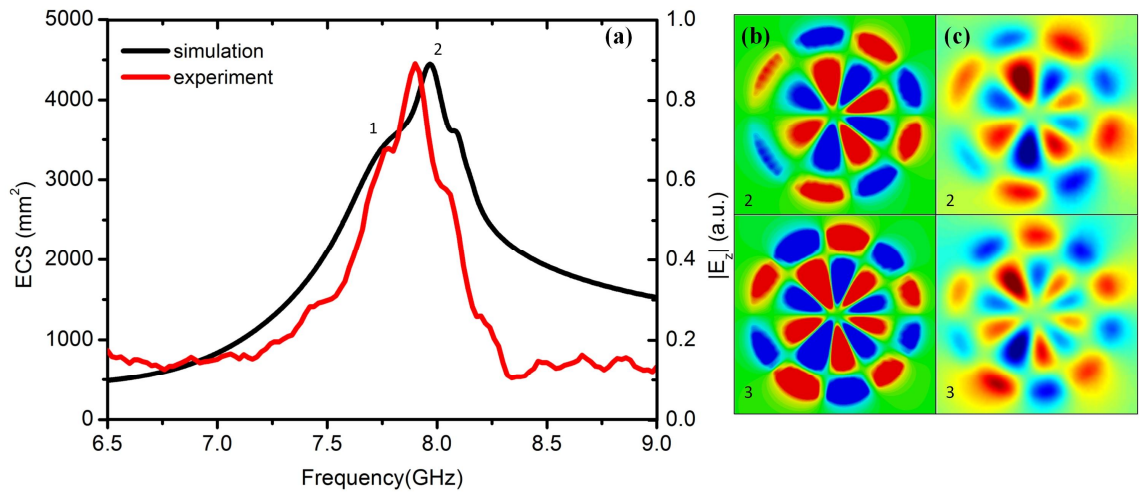

**Figure 8** Simulation and experimental results of the extinction properties (a) and near electric-field characteristics (b and c) of the fabricated ultrathin corrugated disk sample (placed on the top of a thin dielectric substrate, see inset in panel (a). The parameters of the sample are: $R$ =25mm, $r$ =2mm, $L$ =0.018mm, $d$=2.6 mm ( $N$ =60) and $a$=1.3 mm The numerical (b) and experimental (c) near-field maps can be identified with second-order octopole (top) and decapole (bottom) spoof LSP resonances.

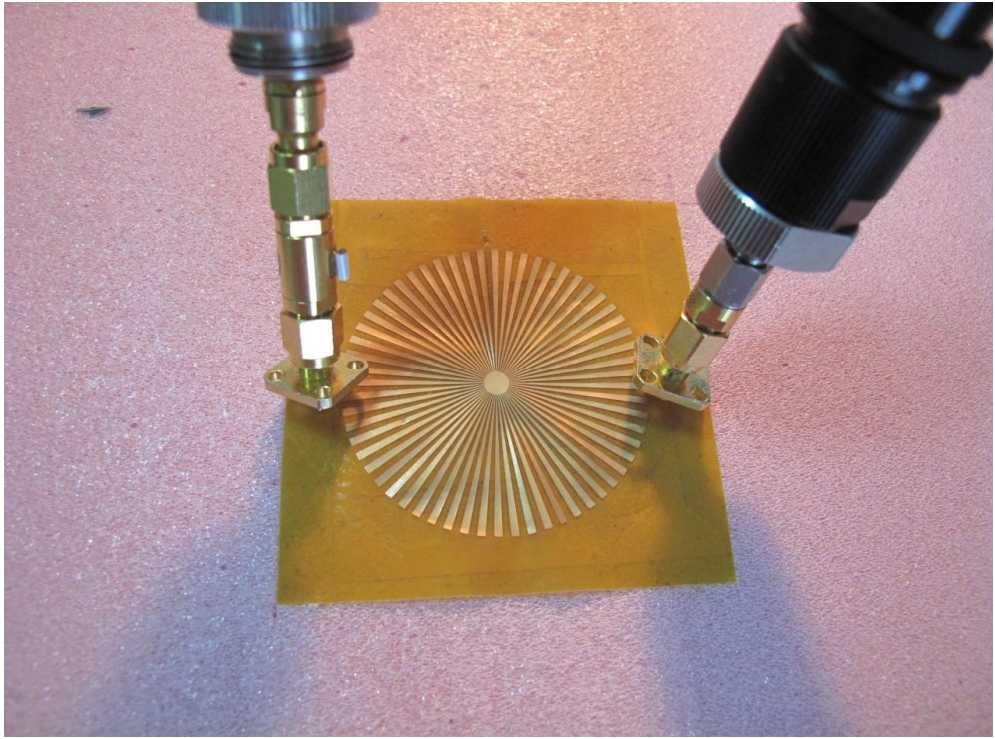

**Figure 9** Photograph of the experimental set up used to measure the near-field characteristics of the spoof LSPs supported by an ultrathin corrugated cooper disk.

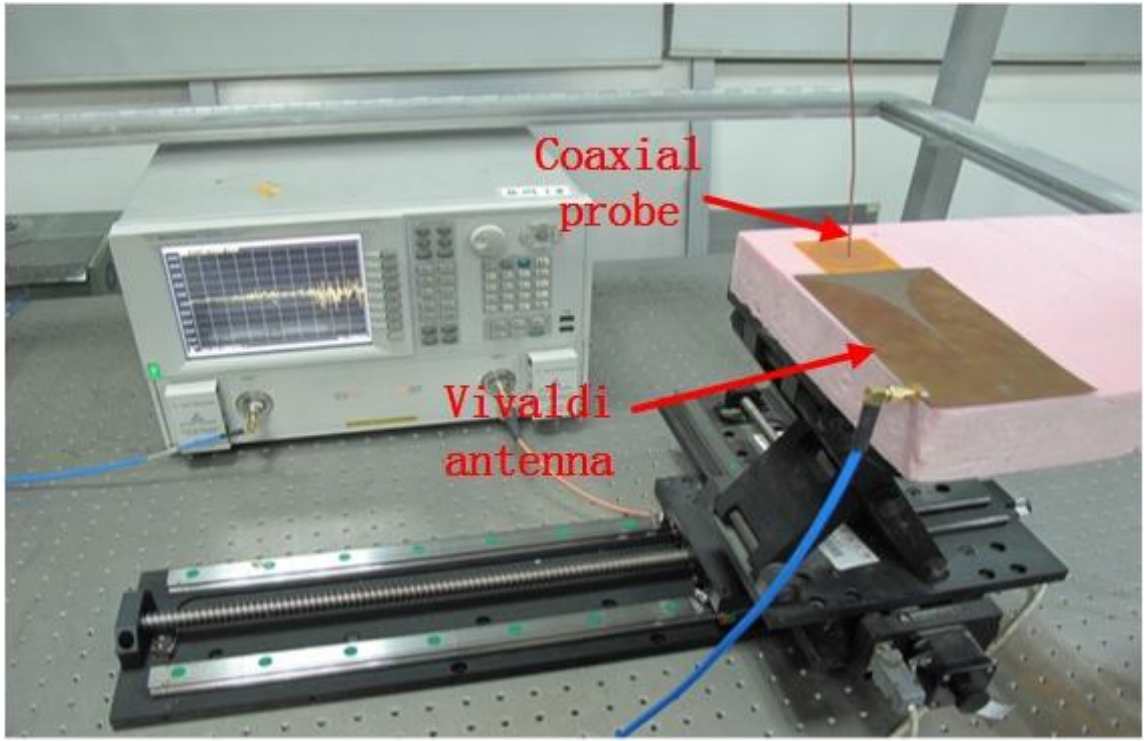

**Figure 10** Photograph of the experimental platform used to measure the spoof LSP near electric-field distributions. The set up consists in a metamaterial Vivaldi antenna as the source, a vector network analyser, a monopole antenna with a 0.2-mm-diameter inner conductor as the detector, and a motion controller.